\documentclass[%
aip,
jmp,
amsmath,amssymb,
reprint,
]{revtex4-1}

\usepackage{graphicx}
\usepackage{dcolumn}
\usepackage{bm}
\usepackage[usenames]{color}

\begin{document}



\title{A comparative study of the low energy HD+$o$-/$p$-H$_2$ rotational excitation/de-excitation
collisions and elastic scattering}


\author{Renat A. Sultanov$^{1,2}$}
\email{rasultanov@stcloudstate.edu}
\affiliation{$^{1)}$Instituto de F\'isica Te\'orica, UNESP $-$ Universidade Estadual Paulista,
01140 S\~ao Paulo, S\~ao Paulo, Brazil\\
$^{2)}$Department of Information Systems \& BCRL, St. Cloud State University,
St. Cloud, Minnesota 56301-4498, USA}





\author{Dennis Guster$^{2}$}
\email{dcguster@stcloudstate.edu}

\author{S. K. Adhikari$^{1}$} 
\email{adhikari@ift.unesp.br}
\homepage{http://www.ift.unesp.br/users/adhikari}




\date{\today}

\begin{abstract}
The Diep and Johnson (DJ) H$_2$-H$_2$ potential energy surface (PES) 
obtained from the first principles
[P. Diep, K. Johnson, J. Chem. Phys. 113, 3480 (2000); 114, 222 (2000)],
has been adjusted through appropriate rotation of the three-dimensional 
coordinate  system  and applied to low-temperature 
($T<300$ K) HD+$o$-/$p$-H$_2$ collisions of astrophysical interest.
A non-reactive quantum mechanical close-coupling method is used
to carry out the computation for the total rotational state-to-state cross sections
$\sigma_{j_1j_2\rightarrow j'_1j'_2}(\epsilon)$
and corresponding thermal rate coefficients
$k_{j_1j_2\rightarrow j'_1j'_2}(T)$.
A rather satisfactory agreement has been obtained between our results computed with the modified
DJ PES and with the newer H$_4$ PES [A.I. Boothroyd, P.G. Martin, W.J. Keogh,
M.J. Peterson, J. Chem. Phys. 116, 666 (2002)], which is also applied in this work.
A comparative study with previous results is presented and discussed.
Significant differences have been obtained for few specific rotational transitions in the
H$_2$/HD molecules  between
our results and previous calculations. The low temperature data for
$k_{j_1j_2\rightarrow j'_1j'_2}(T)$ calculated in this work
can be used in a future application such as a new
computation of the HD cooling function of primordial gas, which is important in the astrophysics of the early Universe.
%
%
%
%
\end{abstract}


\maketitle


\section{Introduction} The simplest quantum-mechanical scattering 
problem, that between two hydrogen molecules, has been a formidable 
challenge to chemists and physicists working in this area. This is 
because the elastic and inelastic H$_2+$H$_2$ scattering have all the 
complications of a complex molecular scattering process like rotational 
and vibrational excitations and de-excitations. Many theoretical and 
experimental methods have been developed and used for the study of these 
processes 
\cite{dj,zarur74,green75,johnson79,schaefer,flower99a,flower00a,flower00b,clary02,guo03,otto5,my1,lee,belof,karl2010}.
Additionally, there has been great interest in the study of 
HD$+$H$_2$ scattering. Although, this system bears some similarity with 
H$_2+$H$_2$, the identical system symmetry is broken here and hence many 
aspects of the theoretical formulation can be tested in the system under 
different symmetry conditions 
\cite{johnson79,chu,buck82,chandler86,chandler88,flower98,flower99,flower99r}.

The PESs of H$_2$+H$_2$ and HD-H$_2$ are basically the same six-dimensional 
functions. The fact follows from a general theoretical point of view and 
the Born-Oppenheimer model\cite{BO}. 
However, from an intuitive physical point of view, the two 
collisions should have rather different scattering output. This is 
because the H$_2$ and HD molecules have fairly different rotational 
constants and rotational-vibrational spectrum. They have different internal 
symmetries and dipole moments. Collisional properties of these systems 
are expected to be highly sensitive to dipole moments.
Further, the HD-H$_2$ PES can be 
derived from H$_2$+H$_2$ by adjusting the coordinate of 
the center of mass of the HD molecule. 
Once the symmetry is broken in H$_2$+H$_2$ by replacing the 
H by the D atom in one H$_2$ molecule one can obtain the HD-H$_2$ PES. 
The new potential has all parts of the full HD-H$_2$ interaction
including HD's dipole moment.
However, in a case when a PES is formulated for fixed interatomic coordinates between the two
hydrogen atoms in the H$_2$-H$_2$ system, it would be difficult to extract the PES of HD-H$_2$, because
the position of the HD center of mass is different from that of H$_2$.
In this work 
we apply a rotational method       for calculation of the HD-H$_2$ PES from that of H$_2$-H$_2$.
The method is based on a rotation of the three-dimensional 
space from the body-fixed H$_2$-H$_2$ coordinate system to appropriately 
adjust the HD molecule center of mass.

The hydrogen molecule plays an important role in many areas of astrophysics. 
For example, the interstellar medium (ISM) cooling process
is associated with the energy loss after inelastic collisions of the particles.
These processes convert the
kinetic energy of ISM's particles to their internal energies: for
instance, in the case of molecules to their internal energy of
rotational-vibrational degrees of freedom. Therefore,
in order to accurately model the thermal balance and 
kinetics of  ISM one needs accurate 
state-to-state cross sections and thermal rate constants.
Theoretical state-resolved treatment requires precise PES of 
{H$_2$-H$_2$}, and a reliable dynamical method for  H$_2$+H$_2$/HD collision 
\cite{chu,schaefer,flower99}. However, on the other hand, experimental measurements of these cross sections 
is a very difficult technical problem. Unfortunately, up to now no reliable experiments are available
on these collision processes, which have important astrophysical applications.
Moreover, different calculations
with various {H$_2$-H$_2$} PESs showed rather different results for important
H$_2$+H$_2$ thermal rate coefficients.

The possible importance of the HD cooling in ISM was first noted in 1972 by
Dalgarno and McCray \cite{dalg}. Because of the special properties of HD
this molecule is even more effective cooler than H$_2$. 
This happens at low temperatures 
$T < 100$ K \cite{dalg},  
where the cooling function of HD becomes $\sim$15 
times larger than the H$_2$ cooling function \cite{dalg}. As we mentioned in the
previous paragraph, to carry out calculation of the
cooling function one needs to know precise cross sections and thermal rate coefficients of the
rotational-vibrational energy transfer collisions. That is why
there is a constant interest to
reliable quantum-mechanical computation of different rotational-vibrational
atom-molecular energy transfer
collision cross sections and corresponding thermal rate
coefficients\cite{roudnev,forrey,bohn1,petrov,bohn2}.


A realistic full-dimensional {\it ab initio} PES for the H$_2$-H$_2$
system was first constructed by Schwenke  \cite{schwenke} and 
that potential was widely used in a variety of methods and 
computation techniques. Flower \cite{flower99}
used Schwenke's H$_2$-H$_2$ PES \cite{schwenke}
in a study of H$_2$-HD collision.
Later on, new extensive studies of the H$_2$-H$_2$ system by 
Diep and Johnson (DJ) \cite{dj} and Boothroyd {\it et al.} (BMKP) \cite{booth}
have produced refined PES for the H$_2$-H$_2$
system. These  PESs have been used in several different calculations
\cite{guo02,montero05,my11,otto8,otto9}.
However, in our previous works \cite{my1,my11} we found that in the case of 
low energy H$_2$+H$_2$ collision
these PESs provide different results for some specific 
state-resolved cross sections.
The difference may be up to an order of magnitude. 
This fact  was also confirmed in work\cite{lee}.
The BMKP PES, probably, needs future improvements {\cite{my1,my11,lee}}, because the
DJ PES gives better agreement with existing experiments for 
H$_2$+H$_2$ than the BMKP potential.

Nonetheless, as a first trial Sultanov {\it et al.} 
applied the BMKP PES to the low-energy 
HD+H$_2$ collisions { 
after appropriate modification \cite{my2,my22}}.
When the results of these calculations  \cite{my2,my22} were 
compared  with prior 
studies
 \cite{schaefer,flower99} 
a relatively good agreement was found
with Schaefer's results \cite{schaefer}. 
At the same time
 substantial differences with the newer data obtained by 
Flower was noted \cite{flower99}.
Therefore, in the light of these circumstances,  it 
would be  useful to apply
another modern H$_2$-H$_2$ PES\cite{dj} to HD-H$_2$ collisions. 
The DJ potential  was
formulated specifically for the symmetric H$_2$-H$_2$ system 
when distances between hydrogen atoms
are fixed at a specific equilibrium value in each H$_2$ molecule.
In this paper we provide the first calculation describing collisions of 
rotationally excited H$_2$ 
and HD molecules using the DJ PES.
We also provide a comparison with our previous calculations\cite {my2,my22} with the BMKP PES,
which is a global six-dimensional surface.

In Sec. \ref{II} we  briefly outline the quantum-mechanical 
close-coupling approach and
our method to convert the symmetric DJ PES to be appropriate for 
calculations of the HD+H$_2$ system.
We represent results for selected state-to-state cross sections and 
thermal rate coefficients
for rotational excitation/de-excitation of HD in low-energy 
collisions with $o$-/$p$-H$_2$ in Sec. \ref{III}. Finally, in Sec. \ref{IV}
we present a brief summary of our findings and conclusion.


\section{Computational method}
\label{II}

\subsection{Dynamical equations}

\begin{figure}
\includegraphics[width=.92\linewidth,clip]{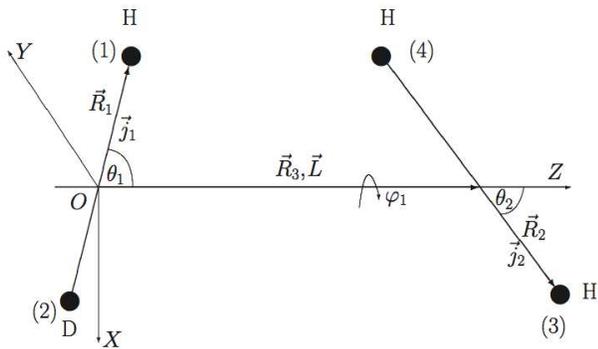}
\caption{(Color online) Four-atomic system $(12)+(34)$ or HD+H$_2$, where H is a hydrogen atom
and D is deuterium,
represented by few-body Jacobi coordinates: $\vec R_1$, $\vec R_2$, and $\vec R_3$.
The vector $\vec R_3$ connects the center of masses of the HD and H$_2$ molecules and
is directed over the axis $OZ$, $\theta_1$ is the angle between $\vec R_1$ and $\vec R_3$,
$\theta_2$ is the angle between $\vec R_2$ and $\vec R_3$, $\varphi_1$ is
the torsional angle,
$j_1,\ \ j_2,\ \ L $ are quantum angular momenta over the corresponding
Jacobi coordinates $\vec R_1$, $\vec R_2$, and $\vec R_3$.  
}
\label{fig1}
\end{figure}

The Schr\"odinger equation for the  $(12)+(34)$ collision in the 
center-of-mass frame,
where {1, 2, 3} and {4} are atoms and $(12)$ and $(34)$ are 
diatomic molecules modeled by 
linear 
rigid rotors, is \cite{green75,green77,miller76}
\begin{eqnarray}
\biggr[\frac{P^2_{\vec R_3}}{2M_{12}}&+&\frac{L^2_{\hat R_{1}}}{2\mu_1R_1^2}+
\frac{L^2_{\hat R_{2}}}{2\mu_2R_2^2}+
V(\vec R_1,\vec R_2,\vec R_3)\nonumber \\ &-& E \biggr]
\Psi(\hat R_{1},\hat R_{2},\vec R_3)=0,
\label{eq:schred}
\end{eqnarray}
where $P_{\vec R_3}$ is the momentum operator of the kinetic energy of 
collision, $\vec R_3$ is the collision 
coordinate, whereas $\vec R_1$ and $\vec R_2$ are relative vectors between 
atoms in the
two diatomic molecules as shown in the Fig. \ref{fig1}, 
and $L_{\hat R_{1(2)}}$ are the quantum-mechanical
rotation operators of the rigid rotors.
Here, $M_{12}\equiv [(m_1+m_2)(m_3+m_4)]/[(m_1+m_2+m_3+m_4)]$ is the 
 reduced mass of the two diatomic molecules $(12)$ and $(34)$ and 
$\mu_{1(2)}\equiv [m_{1(3)}m_{2(4)}]/[(m_{1(3)}+m_{2(4)})]$ are 
reduced masses of the two molecules. The vectors
$\hat R_{1(2)}$ are the angles of 
orientation of rotors $(12)$ and $(34)$, respectively;
$V(\vec R_1,\vec R_2,\vec R_3)$ is the PES for the four-atomic 
system $(1234)$, and 
$E$ is the total energy in the center-of-mass system. 
The linear rigid-rotor model used in this calculation 
was already applied in some previous
studies \cite{flower99,green77,miller76}.
For the considered range of kinetic energies of astrophysical interest
the model is quite justified.

The eigenfunctions of the operators $L_{\hat R_{1(2)}}$ 
in Eq. (\ref{eq:schred}) are simple spherical harmonics 
$Y_{j_im_i}(\hat r)$. To solve this equation
the following angular-momentum expansion is used:
\begin{eqnarray}
\Psi(\hat R_{1},\hat R_{2},\vec R_3)&=&\sum_{JMj_1j_2j_{12}L} 
\frac{U^{JM}_{j_1j_2j_{12}L}(R_3)}{R_3}\nonumber \\&
\times&\phi^{JM}_{j_1j_2j_{12}L}(\hat R_1,\hat R_2,\vec R_3)
\label{eq:expn}
\end{eqnarray}
where $U^{JM}_{j_1j_2j_{12}L}(R_3)$ are unknown coordinate functions, 
$J$ is total angular momentum quantum number of the 
$(1234)$ system and $M$ is its projection 
onto the space fixed $Z$ axis, and channel expansion functions are the 
following:
\begin{eqnarray}
&&\phi^{JM}_{j_1j_2j_{12}L} (\hat R_1,\hat R_2,\vec R_3)=
\sum_{\tilde m_1 \tilde m_2 \tilde m_{12} \tilde m}
C_{j_1 \tilde m_1 j_2 \tilde m_2}^{j_{12} \tilde m_{12}}
C_{j_{12} \tilde m_{12}l \tilde m}^{JM}
\nonumber \\
&\times& 
Y_{j_1 \tilde m_1}(\hat R_1)
Y_{j_2 \tilde m_2}(\hat R_2)Y_{L \tilde m}(\hat R_3),
\end{eqnarray}
here $j_1+j_2=j_{12}$, $j_{12}+L=J$, $\tilde m_1$, $\tilde m_2$, $\tilde m_{12}$ and $\tilde m$ are projections
of $j_1$, $j_2$, $j_{12}$ and $L$, respectively.
{The $C$'s are the appropriate Clebsch-Gordan coefficients.}
The quantum-mechanical momenta $j_1$, $j_2$ and $L$ over corresponding 
Jacobi vectors 
$\vec R_1$, $\vec R_2$ and $\vec R_3$ are also shown in Fig.\ 1.

Upon
substitution of Eq. (\ref{eq:expn}) into Eq. (\ref{eq:schred}), one obtains  
 a set of coupled
second order differential equations for the unknown radial functions 
$U^{JM}_{\alpha}(R_3)$ \cite{green77,miller76}
\begin{eqnarray}
&&\left(\frac{d^2}{dR_3^2}-\frac{L(L+1)}{R_3^2}+k_{\alpha}^2\right)U_{\alpha}^{JM}(R_3)
\nonumber \\&&
=2M_{12}
\sum_{\alpha'} \int <\phi^{JM}_{\alpha}(\hat R_1,\hat R_2,\vec R_3)
|V(\vec R_1,\vec R_2,\vec R_3)|
\nonumber \\&&
\phi^{JM}_{\alpha'}(\hat R_1,\hat R_2,\vec R_3)>U_{\alpha'}^{JM}(R_3) d\hat R_1 d\hat R_2 d\hat R_3,
\label{eq:cpld}
\end{eqnarray}
where $\alpha \equiv (j_1j_2j_{12}L)$.
We apply the hybrid modified log-derivative-Airy propagator in the 
general-purpose scattering program MOLSCAT \cite{hutson94} to solve the 
coupled radial Eq. (\ref{eq:cpld}).

The log-derivative matrix is propagated to large intermolecular 
distances $R_3$, since all experimentally observable
quantum information about the collision is contained in the asymptotic 
behavior of functions 
$U^{JM}_{\alpha}(R_3\rightarrow\infty)$. The numerical results are matched 
to the known 
asymptotic solution to derive the  scattering $S$-matrix 
$S^J_{\alpha \alpha'}$:
\begin{eqnarray}
U_{\alpha}^J
&\mathop{\mbox{\large$\sim$}}\limits_{R_3 \rightarrow + \infty}
&\delta_{\alpha \alpha'}
e^{-i(k_{\alpha \alpha}R_3 - l\pi/2)}
- \sqrt{\frac{k_{\alpha \alpha}} {k_{\alpha \alpha'}}} S^J_{\alpha \alpha'}\nonumber \\
&\times & e^{-i(k_{\alpha \alpha'}R_3 - l'\pi/2)},
\end{eqnarray}
where $k_{\alpha \alpha'}=\sqrt{2M_{12}(E+E_{\alpha}-E_{\alpha'})}$
is the channel wave number, $E_{\alpha(\alpha')}$
are rotational channel energies for $\alpha$ and $\alpha '$.
The method was used for each partial wave until a converged cross section was obtained. 
It was verified that the results are converged with respect to the number of partial waves as well as
the matching radius, $R_{3max}$, for all channels included in our calculation.
The cross sections for rotational excitation and relaxation can 
be obtained directly from the $S$-matrix.
In particular the cross sections for excitation from $j_1j_2\rightarrow j'_1j'_2$
summed over the final $\tilde m'_1 \tilde m'_2$ and averaged over the initial 
$\tilde m_1 \tilde m_2$ are given by
\begin{eqnarray}
\sigma(j'_1,j'_2;j_1j_2,\epsilon) = \pi \sum_{Jj_{12}j'_{12}LL'}(2J+1)\nonumber \\
\times \frac{|\delta_{\alpha\alpha'} - S^J(j'_1,j'_2,j'_{12}L';j_1,j_2,j_{12},L; E)|^2}
{(2j_1+1)(2j_2+1)k_{\alpha\alpha'}}
\label{eq:sigma}
\end{eqnarray}
The relative
kinetic energy of the two molecules in the center of mass frame is: 
\begin{equation}
\epsilon=E-B_1j_1(j_1+1)-B_2j_2(j_2+1),
\label{eq:ekin}
\end{equation}
where $B_{1(2)}$ are the rotation constants of rigid rotors $(12)$ and 
$(34)$, respectively, of total angular momentum $j_{1(2)}$.
The relationship between the rotational thermal-rate coefficient 
$k_{j_1j_2\rightarrow j'_1j'_2}(T)$ at temperature $T$
and the corresponding
cross section $\sigma_{j_1j_2\rightarrow j'_1j'_2}(\epsilon)$, 
 can be obtained 
through the following
weighted average \cite{billing}
\begin{eqnarray}
k_{j_1j_2\rightarrow j'_1j'_2}(T) &=& \frac{1}{(k_BT)^2} \sqrt{ \frac{8k_B T}{\pi M_{12}}}
\int_{\epsilon_s}^{\infty}
\sigma_{j_1j_2\rightarrow j'_1j'_2}(\epsilon)\nonumber \\
&\times & e^{-\epsilon/k_BT}\epsilon d\epsilon,
\label{eq:kT}
\end{eqnarray}
where 
$k_B$ is Boltzman constant  and 
$\epsilon_s$ is the minimum relative kinetic energy of the two molecules 
for the levels $j_1$ and $j_2$
to become accessible.

\subsection{The  BMKP   H$_2$-H$_2$ PES}

\label{IIB}

The BMKP PES \cite{booth},
is a global six-dimensional potential energy surface for two hydrogen molecules.
It was especially constructed to represent the whole interaction region of the chemical reaction dynamics
of the  four-atomic system and to provide an accurate  van der 
Waals well. In the six-dimensional conformation space of the four-atomic system the PES
forms a complicated three-dimensional hyper surface \cite{booth}.
To compute the distances between the four atoms { 
$R_1,R_2$ and $R_3$},
the BMKP PES uses Cartesian coordinates.
Therefore it was necessary to convert spherical coordinates used in
the close-coupling method \cite{hutson94} to the corresponding Cartesian 
coordinates and compute
the distances between the four atoms followed by calculation of 
the PES \cite{my2,my22}.
Without the loss of generality the procedure used a specifically 
oriented coordinate system $OXYZ$ \cite{my2}.
First we introduce the Jacobi coordinates $\{\vec R_1, \vec R_2, \vec R_3\}$ and the radius-vectors 
of all four atoms in the space-fixed coordinate system 
$OXYZ$: $\{\vec r_1, \vec r_2, \vec r_3, \vec r_4\}$ (not shown in Fig. \ref{fig1}).
As in Ref.~\onlinecite{my2} we apply the following procedure: 
the center of mass of the HD molecule is put at 
the origin of the coordinate system $OXYZ$,
and the $\vec R_3$ is directed to center of mass of 
the H$_2$ molecule along the $OZ$ axis, as shown in Fig. \ref{fig1}.
Then $\vec R_3=\{R_3, \Theta_3=0, \Phi_3=0\}$, with $\Theta_3$ and 
$\Phi_3$ the polar and azimuthal angles,
$\vec R_1=\vec r_1 - \vec r_2$, $\vec R_2=\vec r_4 - \vec r_3$,
$\vec r_1=\xi \vec R_1$ and $\vec r_2=(1 - \xi)
 \vec R_1$, where $\xi=m_2/(m_1+m_2)$.

Next, without the loss of generality, we can adopt the $OXYZ$ system in such a way, that
the HD inter-atomic vector $\vec R_1$ lies on the $XOZ$ plane. Then the angle variables of
$\vec R_1$ and $\vec R_2$ are: $\hat R_1=\{\Theta_1,\Phi_1=\pi\}$
and $\hat R_2=\{\Theta_2,\Phi_2\}$ respectively.
One can see, that the Cartesian coordinates of the atoms of the HD molecule are:
\begin{eqnarray}
\vec r_1 &=& \{x_1=\xi R_1\sin\Theta_1, y_1=0, z_1=\xi R_1 \cos\Theta_1\},\\
\vec r_2 &=& \{x_2=-(1-\xi) R_1 \sin\Theta_1, y_2=0, \nonumber \\
z_2&=& -(1-\xi) R_1 \cos\Theta_1\}.
\end{eqnarray}
{ Defining  $\zeta = m_4/(m_3+m_4)$, we have} 
\begin{eqnarray}
\vec r_3 = \vec R_3 - (1-\zeta)\vec R_2,\\
\vec r_4 = \vec R_3 + \zeta \vec R_2,
\end{eqnarray}
{and the corresponding Cartesian coordinates:} 
\begin{eqnarray}
\vec r_3 = \{x_3&=&-(1-\zeta) R_2 \sin\Theta_2 \cos\Phi_2,\nonumber \\ 
y_3&=&-(1-\zeta)R_2\sin\Theta_2\sin\Phi_2,\nonumber \\
z_3&=& R_3-(1-\zeta)R_2\cos\Theta_2\},
\end{eqnarray}
\begin{eqnarray}
\vec r_4 = \{x_4&=&\zeta R_2 \sin\Theta_2 \cos\Phi_2, \nonumber \\
y_4&=&\zeta R_2 \sin\Theta_2\sin\Phi_2,\nonumber \\
z_4&=& R_3+\zeta R_2\cos\Theta_2\}.
\end{eqnarray}
%

\begin{figure}
\includegraphics[width=.92\linewidth,clip]{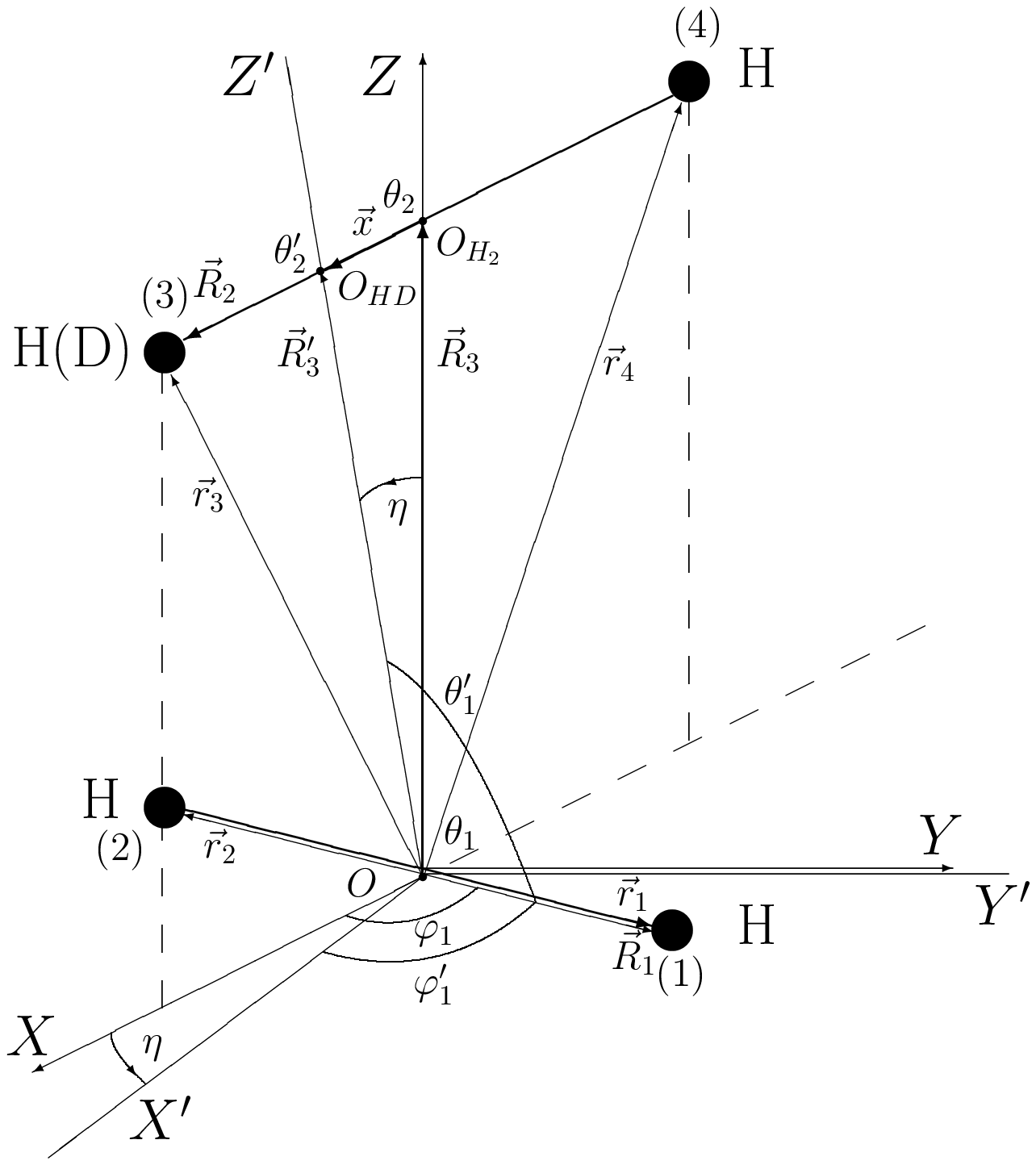}
\caption{(Color online)  The geometrical configuration of the
HD+H$_2$ system in the framework of two different Cartesian coordinate systems:
$OXYZ$ and $OX'Y'Z'$. Here: $\vec x$ is the distance between $O_{H_2}$ and $O_{HD}$,
$\eta$ is the rotation angle over the $OZ$ axis, i.e. the angle between the axes $OZ$
and $OZ'$ and
between the axes $OX$ and $OX'$ , $\vec r_1,\ \vec r_2,\ \vec r_3$ and $\vec r_4$ are the
radius-vectors of the
corresponding atoms. The vectors $\vec R_2$,  $\vec R_3$, and
$\vec R'_3$, belong to
the $XOZ$ plain.}
\label{fig2}
\end{figure}

\subsection{The Modified Diep and Johnson H$_2$-H$_2$ potential}

\label{IIC}

Below we briefly present our method to convert the symmetric DJ H$_2$-H$_2$ PES \cite{dj}
to be suitable for the non-symmetric system HD+H$_2$. The method is based on a mathematical
transformation technique, i.e. a geometrical rotation of the three-dimensional 
(3D) space and the corresponding space-fixed coordinate system $OXYZ$\cite{NASA2010}, as 
shown in Fig. \ref{fig2}. It is more convenient here to use a slightly different from Fig. \ref{fig1} orientation 
of the coordinate system $OXYZ$: now the center of mass of the H$_2$ molecule is set at the origin  
of the space-fixed $OXYZ$, and $\vec R_3$  $(\vec R'_3)$
is directed to the center of mass of H$_2$ (HD).

The few-body system (1234)   
can be characterized by four radius-vectors:
$\{\vec r_1, \vec r_2, \vec r_3, \vec r_4\}$, or alternatively
by so-called three Jacobi vectors: $\{\vec R_1, \vec R_2, \vec R_3\}$.
Usually the second option is more convenient for describing
the quantum-mechanical few-body systems.
Next, the initial geometry of the system is taken in such a way that 
the Jacobi vector $\vec R_3$ connects the center of masses of the two 
molecules and is
directed over the ${OZ}$ axis. We can also choose $OXYZ$ in such a manner 
that the Jacobi vector $\vec R_2$ lies in the $X$-$Z$ plane.  
Finally, the vector $\vec R_1$ can be directed anywhere. 
Then the spherical coordinates of the
Jacobi vectors are: $\vec R_1=(R_1, \theta_1, \phi_{12})$, $\vec R_2=(R_2, \theta_2, 0)$, and
$\vec R_3=(R_3, 0, 0)$. Thus the 4-body system H$_2$-H$_2$/HD can be fully determined
with the use of six variables. However, the DJ PES has been prepared
for the rigid monomer model of H$_2$-H$_2$ \cite{dj}, 
so actually we have only four active variables in this consideration:
$R_3, \theta_1, \theta_2$, and $\phi_{12}$.

The DJ PES \cite{dj} has been prepared as the following function for the distance $R_3$ between two
H$_2$ molecules, two polar angles $\theta_{1(2)}$ and one torsional angle $\phi_{12}$:
\begin{equation}
V(R_3, \theta_1, \theta_2, \phi_{12}) = \sum_{l_1,l_2,l} V_{l_1,l_2,l}(R_3)
G_{l_1,l_2,l}(\theta_1, \theta_2, \phi_{12}),
\label{eq:dj}
\end{equation}
where the angular functions $G_{l_1,l_2,l}$ are \cite{dj}:
\begin{eqnarray}
&&G_{000}(\theta_1, \theta_2, \phi_{12}) = 1, \\ 
&&G_{202}(\theta_1, \theta_2, \phi_{12}) = \frac{5}{2}(3\cos^2\theta_1-1), \\
&&G_{022}(\theta_1, \theta_2, \phi_{12}) = \frac{5}{2}(3\cos^2\theta_2-1),\\ 
&&G_{224}(\theta_1, \theta_2, \phi_{12}) = \frac{45}{4\sqrt{70}}
[2(3\cos^2\theta_1-1)\nonumber \\&& 
(3\cos^2\theta_2-1)-
16\sin\theta_1\cos\theta_1\sin\theta_2\nonumber \\
&\times&\cos\theta_2\cos\phi_{12}
 + \sin^2\theta_1\sin^2\theta_2\cos(2\phi_{12})].
\label{eq:gll}
\end{eqnarray}
The coordinate function $V_{l_1,l_2,l}(R_3)$ has been tabulated in
work\cite{dj}.

First we start with the original DJ PES
of Eqs.  (\ref{eq:dj})$-$(\ref{eq:gll}).
Then  we replace one hydrogen atom ``H" with a 
deuterium atom ``D". Thus
we break the symmetry by shifting the center of 
mass of one H$_2$ molecule 
to another point, that is from $O_{H_2}$ to $O_{HD}$ as  
shown in Fig. \ref{fig2},
we also need to take into acount the difference between the number of
rotational states in H$_2$+H$_2$ and HD+H$_2$.
The length of the vector $\vec x$  is $x = |\vec R_{2}|/6$.
Now  we rotate the $OXYZ$ coordinate system
around the ${OY}$ axis in such a way that the new ${OZ'}$ axis goes
through the point O$_{HD}$. The ${OY'}$ axis and the old ${OY}$ axis are 
parallel, and
the angle of this small rotation is $\eta$, see Fig. \ref{fig2}.
This transformation converts the initial Jacobi vectors in $OXYZ$:
$\vec R_1 = \{R_1, \theta_1, \phi_{12}\}$, 
$\vec R_2 = \{R_2, \theta_2, 0\}$ and 
$\vec R_3 = \{R_3, 0, 0\}$ to the corresponding Jacobi vectors with new coordinates in the
new $O'X'Y'Z'$: $\vec R'_1 = \{R'_1, \theta'_1, \phi'_{12}\}$,
$\vec R'_2 = \{R'_2, \theta'_2, 0\}$ and $\vec R'_3 = \{R'_3, 0, 0\}$.
As a result of this simple procedure we obtain a new PES, namely:
\begin{equation}
V^{\small{\mbox{H}_2}}_{DJ}(R_3, \theta_1, \theta_2, \phi_{12})\rightarrow
\tilde V^{\small{\mbox{HD}}}_{DJ}(R'_3, \theta_1', \theta_2', \phi_{12}').
\label{eq:transfor}
\end{equation}
It is quite obvious, that
the rotation does not affect the coordinate function $V_{l_1,l_2,l}(R_3)$ in Eq. (\ref{eq:dj}).
Note, 
in a previous calculation of low energy
collision between monodeuterated ammonia NH$_2$D and a helium atom,
a somewhat similar spatial rotational-translational
procedure has been applied to the original PES of the NH$_3$-He system\cite{roueff1}.

Next, any rotation of the 3D $OXYZ$ coordinate system can be represented by Euler angles, i.e.
$\{\alpha, \beta, \gamma\}$\cite{varshal88,eric}.
In this work we choose the following Euler angles: $\{ \alpha=0, \beta=\eta, \gamma=0 \}$. 
To calculate the value of the rotational angle $\eta$ in Fig. \ref{fig2}, one can consider the 
internal triangle $\bigtriangleup O_{HD}OO_{H_2}$ which is shown in Fig. \ref{fig2}.
The angle $\eta$ is determined from the following formula:
\begin{equation}
\cot\ \eta=\frac{(R'_3+x\sin\ \theta'_2)}{x\cos\ \theta'_2}.
\label{eq:eta}
\end{equation}
The derivation of (\ref{eq:eta}) can be expressed as follows. First, the angles of the triangle 
$\bigtriangleup O_{HD}O_{H_2}O$ satisfy the following equation 
$(\pi-\theta_2)\ +\ \eta\ +\ \theta'_2\ =\ \pi$ or $\theta_2\ =\ \eta\ +\ \theta'_2$.
Secondly, based on the law of sines for $\bigtriangleup O_{HD}OO_{H_2}$:
\begin{equation}
\frac{x}{sin\ \eta} = \frac{R}{sin\ \theta_2'} = \frac{R'}{sin\ \varepsilon},
\end{equation}
where $\varepsilon\ =\ \pi\ -\ \theta_2$. 
Because $sin\ \varepsilon = cos\ \theta_2$ we have:
\begin{equation}
\frac{x}{sin\ \eta}=\frac{R}{sin\ \theta_2'}=\frac{R'}{cos\ \theta_2},
\end{equation}
and finally:
\begin{equation}
\frac{cos(\eta\ +\ \theta_2')}{sin\ \eta}=\frac{R'}{x},
\end{equation}
from which one can directly obtain the expression (\ref{eq:eta}).  
In such a way the rotation of the
coordinate system from $OXYZ$ to $O'X'Y'Z'$ makes a corresponding transformation of the
coordinates of the atoms in the 4-body system and the distance between two molecules.
One has the following relations among 
old and new variables \cite{varshal88}:
\begin{eqnarray}\label{eq:RX}
\cos(\theta_1) &=& \cos(\theta_1')\cos(\eta) - 
\sin(\theta_1')\sin(\eta)\cos(\phi_{12}'),\hspace{6mm} \\
\cos(\theta_2) &=& \cos(\theta_2')\cos(\eta) 
- \sin(\theta_2')\sin(\eta)\cos(\phi_{12}'),\hspace{6mm} \\
\cot(\phi_{12}) &=& \cot(\phi_1')\cos(\eta) + \cot(\theta_1')\frac{\sin(\eta)}  
{\sin(\phi_{12}')}, \\
R_3 &=& \sqrt{x^2 + R_3'^2 - 2xR_3'\cos(\theta_2')}.
\label{eq:R3}
\end{eqnarray}
In the calculation of HD+H$_2$ with the DJ PES one has to use new 
coordinates $\theta'_1, \theta'_2, \phi'_{12}, R'_3$. However, the 
potential (\ref{eq:dj}) has been expressed through the old H$_2$-H$_2$ 
variables, hence it is to be transformed to new variables using Eqs. 
(\ref{eq:RX}) $-$
(\ref{eq:R3}).



\section{Numerical Results}
\label{III}

Results for the low energy HD+$o$-/$p$-H$_2$ elastic scattering cross-sections and
few selected quantum-mechanical rotational transitions together with the 
corresponding results from previous studies \cite{my2,my22} are presented below.
The cross sections are also compared with the corresponding results from\cite{schaefer}.
Additionally, we compare our results for the thermal rate
coefficients (\ref{eq:kT})  
with previous calculations \cite{schaefer,flower99}.
The results for the low energy elastic scattering cross sections cannot be
compared with other theoretical/experimental data. To the best of our 
knowledge such calculations do not exist.

\begin{figure}
\begin{center}
\includegraphics[width=.92\linewidth,clip,height=11.5pc]{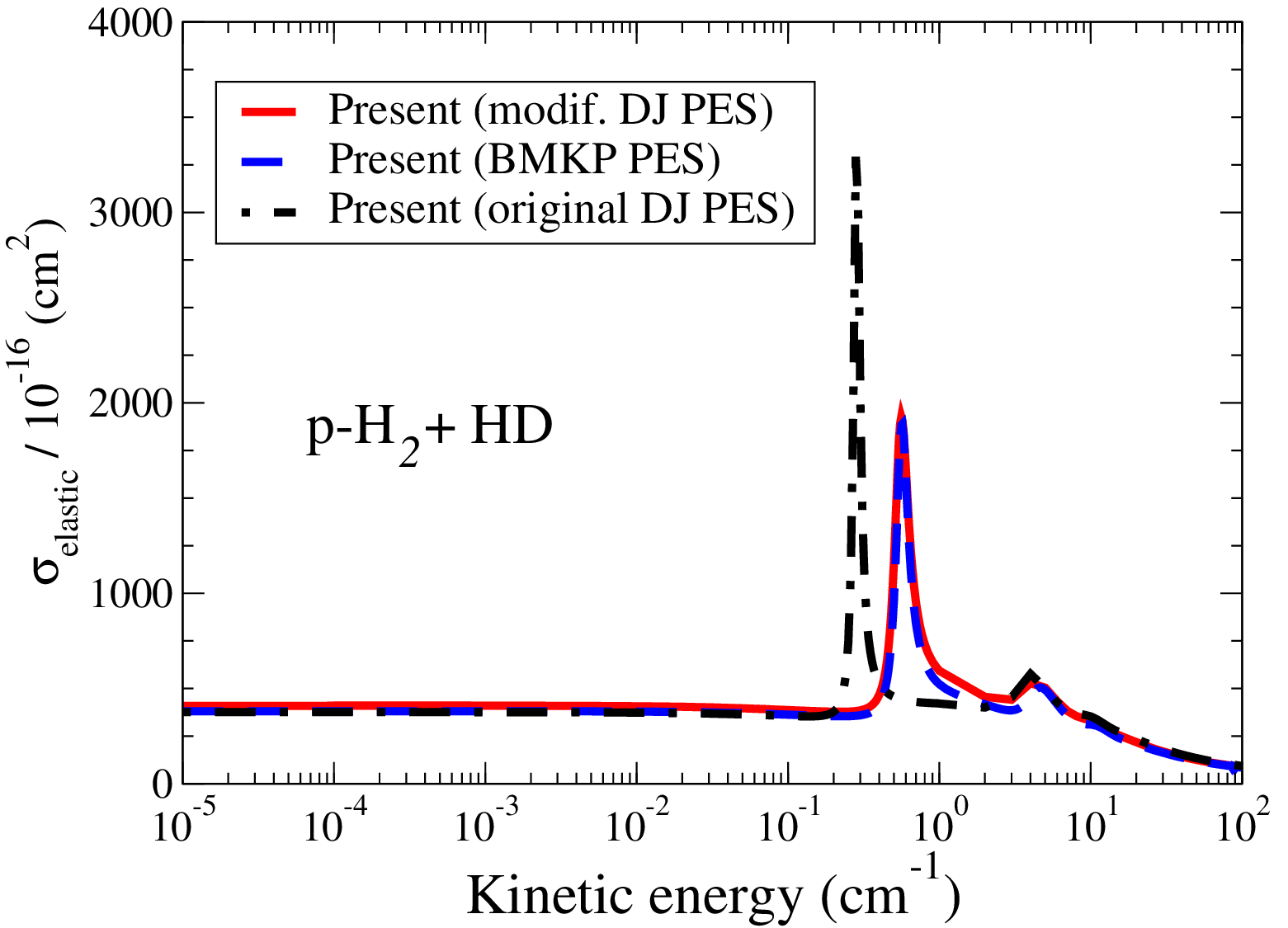}
\includegraphics[width=.92\linewidth,clip,height=11.5pc]{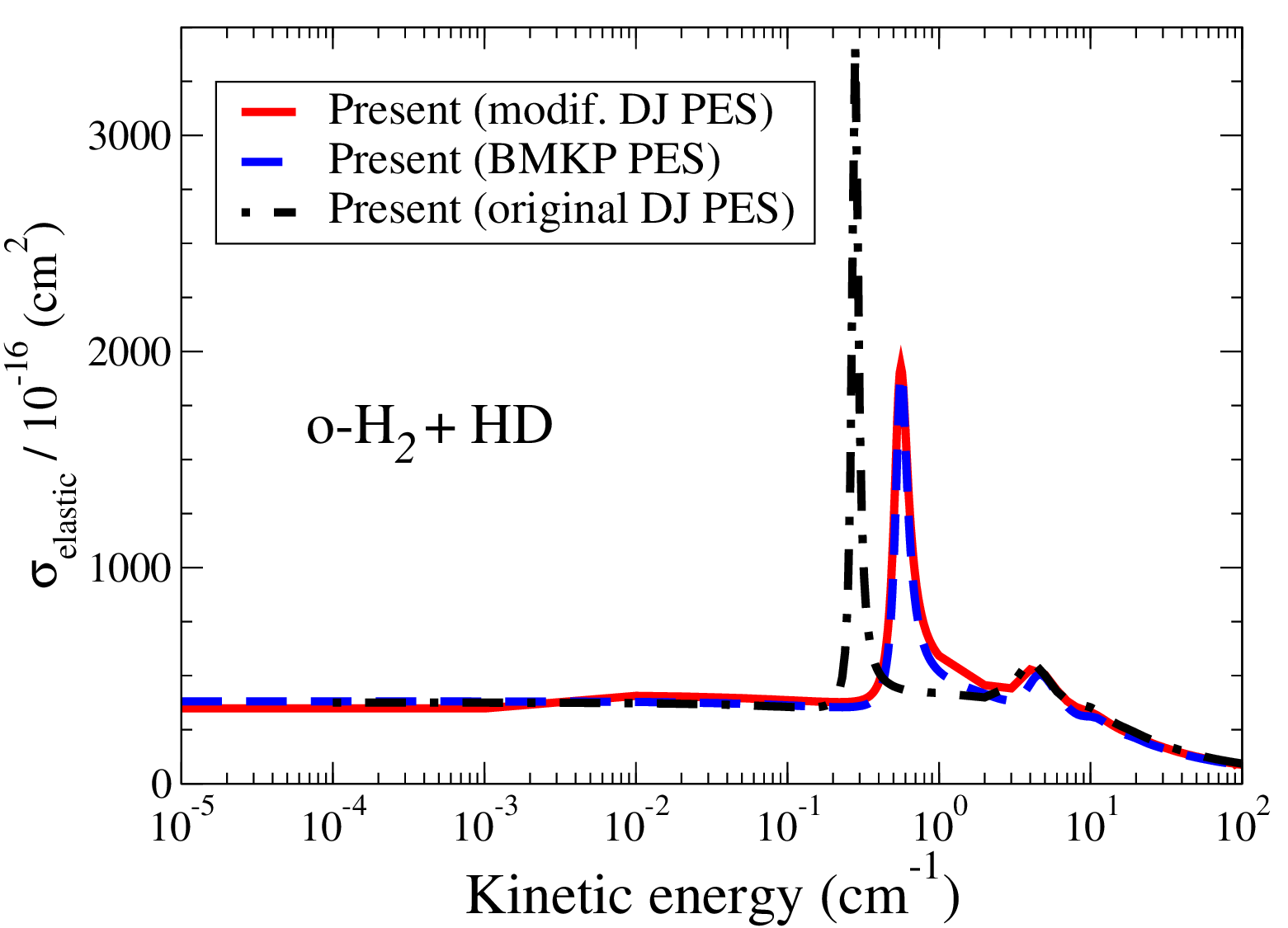}
\caption{(Color online) Elastic scattering total cross sections for 
HD+$o$-H$_2$ ({upper panel}) and HD+$p$-H$_2$ 
({lower panel}) at different 
kinetic energies 
$\epsilon$ with the  BMKP \cite{booth} potential of Sec. 
\ref{IIB}, the modified DJ \cite{dj} PES of Sec. \ref{IIC} and the 
original DJ PES. 
}
\label{fig3} \end{center} \end{figure}

\begin{table*}
\label{I}
\caption{The elastic scattering cross sections $\sigma_{el}$
$(10^{-16}\mbox{cm}^2)$ at selected relative kinetic
 energies $\epsilon$ (cm$^{-1}$)
and corresponding scattering lengths $a_{scatt}$ $(10^{-8}\mbox{cm})$
in the $o$-/$p$-H$_2$ + HD $\rightarrow$ $o$-/$p$-H$_2$ + HD low energy collisions
calculated with three different potentials: modified DJ PES from Sec. \ref{IIC}
and original BMKP \cite{booth} and DJ \cite{dj} PESs.
Numbers in parentheses are powers of 10.}
\vspace{3mm}
\centering
\label{table:1}
\begin{ruledtabular}
\begin{tabular}{lccccccc}
$\epsilon$ (cm$^{-1}$) & \multicolumn{7}{c}{Elastic cross section:
$\sigma_{el}  \times 10^{-16}$ (cm$^2$)}\\
\hline
&\multicolumn{3}{c}{$o$-H$_2$ + HD}&  &\multicolumn{3}{c}{$p$-H$_2$ + HD}\\
\hline
        & mod. DJ & BMKP   & DJ     &  & mod. DJ & BMKP   & DJ    \\
\hline
4.0(-5) & 348.61 & 380.70 & 375.29 &  & 408.68 & 380.02 & 375.87 \\
5.0(-5) & 348.61 & 380.70 & 375.29 &  & 408.68 & 380.02 & 375.87 \\
1.0(-4) & 348.59 & 380.67 & 375.27 &  & 409.96 & 380.00 & 375.86 \\
1.0(-2) & 406.4  & 378.1  & 372.5  &  & 406.6  & 378.5  & 373.0  \\
1.0(-1) & 386.4  & 362.2  & 355.3  &  & 386.6  & 362.5  & 355.8  \\
3.0(-1) & 385.0  & 359.0  & 1705.0 &  & 385.0  & 359.1  & 2039.8 \\
4.0(-1) & 467.3  & 418.9  & 516.26 &  & 466.1  & 416.9  & 529.0  \\
6.0(-1) & 1641.2 & 1577.6 & 433.8  &  & 1653.2 & 1611.5 & 437.0  \\
1.0     & 592.0  & 520.1  & 419.3  &  & 593.3  & 522.6  & 421.0  \\
5.0     & 504.4  & 492.9  & 501.5  &  & 504.4  & 492.8  & 501.8  \\
10      & 331.1  & 310.7  & 353.3  &  & 331.1  & 310.6  & 353.1  \\
100     & 87.6   & 84.6   & 93.2   &  & 87.6   & 84.6   & 93.1   \\
\hline
\hline
& \multicolumn{7}{c}{Scattering length: $a_{scatt}$$\times 10^{-8}$ (cm)} \\
\hline
$\epsilon\rightarrow 0.0$ & 5.27 & 5.50 & 5.46 &  & 5.70 & 5.50 & 5.47 \\
\end{tabular}
\end{ruledtabular}
\end{table*}

\subsection{Elastic scattering}
\label{IIIA}

In Fig. \ref{fig3} we show the present
elastic scattering cross sections for 
HD+$o$-/$p$-H$_2$ collisions at low and ultra-low 
energies calculated using three different potentials:
the {  BMKP potential of Sec. \ref{IIB},
the modified DJ potential of Sec. \ref{IIC}, and the original DJ 
potential appropriate for the H$_2$-H$_2$ system.}
In each cross section there is a prominent resonance.
In Fig. \ref{fig3} although the shapes of three cross sections are  
similar to each other the resonance peak 
and the position of the resonance differ significantly 
when we use the original DJ potential without
the modifications described in the Sec. \ref{II}.
This is because the original, symmetrical DJ potential \cite{dj} does not
have all the asymmetrical features of the HD+H$_2$ interaction. 
Moreover, these features are of crucial 
importance for HD+H$_2$ scattering. In Table \ref{I} we show the 
 elastic cross sections $\sigma_{el}(\epsilon)$  for a few selected  
energies calculated with the three potentials. We also include
in  Table \ref{I} the results for scattering lengths $a_0$'s.
As can be seen from Fig. \ref{fig3} and Table \ref{I}  
we have obtained extremely good agreement between 
our calculations with the BMKP and with the modified DJ PESs. 
For the considered range of the kinetic energies
we obtained full numerical convergence,
for instance, the total angular momentum $J$ was used up to the maximum value $J_{max}=12$
in this calculation.

\subsection{Non-elastic channels}
The main goal of the present study  is not to obtain results for every possible 
transition cross section
in the HD+H$_2$ collision, but rather to demonstrate how the appropriately undertaken
3D rotation of the symmetrical surface of the H$_2$ - H$_2$ system could be adapted for
collision of the H$_2$ and HD molecules. 
We carry out calculations for  a fairly
wide region of the collision energies, i.e. from 3 K to up to 300 K. 
This temperature {interval} 
is relevant for future calculations of the important
HD-cooling function \cite{dalg}.  
Here
we compute few transition cross sections
in which we noticed substantial differences between our results obtained
with the two different PESs from Refs.~\onlinecite{dj,booth}
and also between our results and the data from 
previous studies \cite{schaefer,flower99}.

\begin{figure}
\begin{center}
\includegraphics[width=.92\linewidth,clip,height=11.5pc]{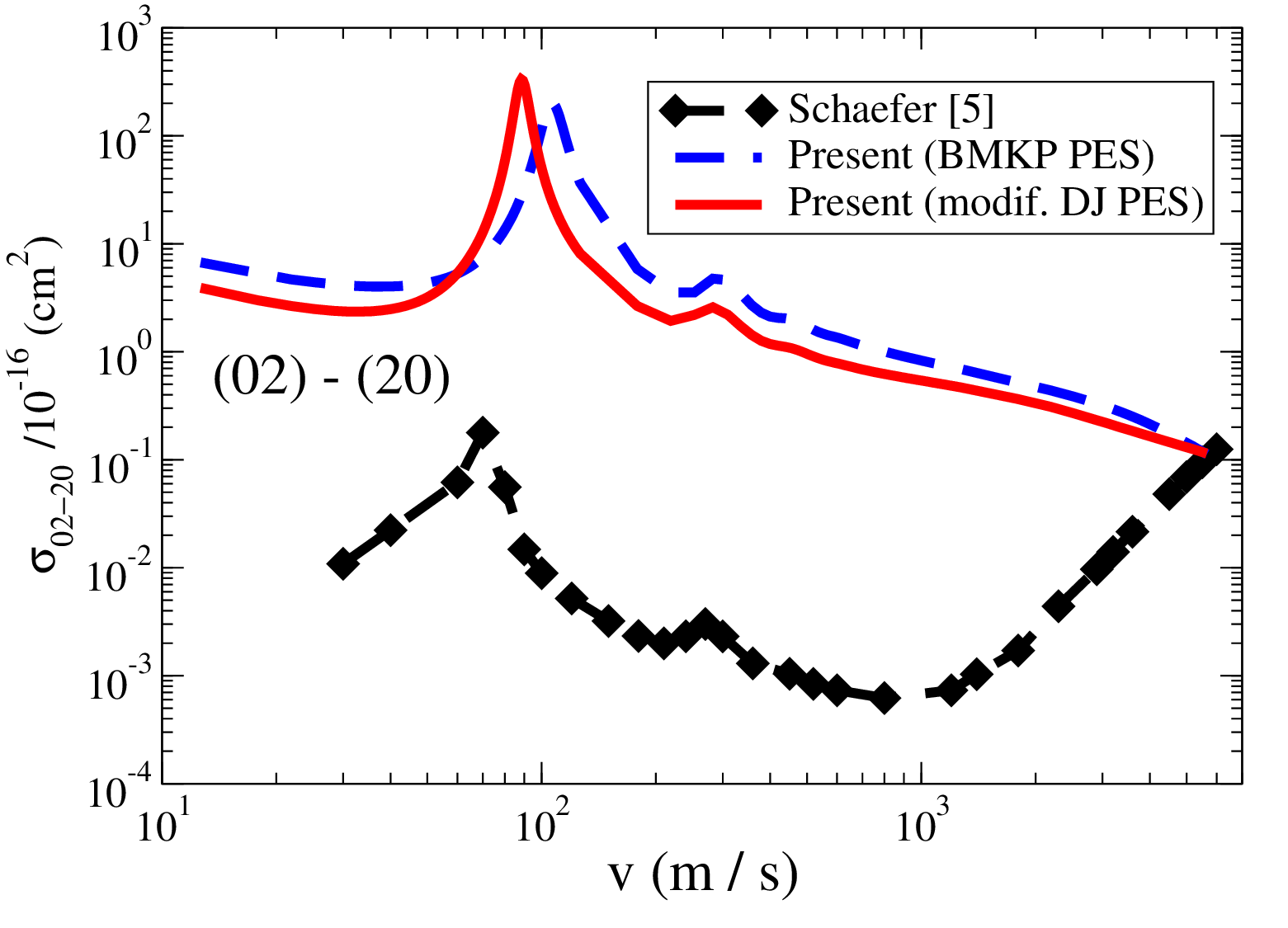}   
\includegraphics[width=.92\linewidth,clip,height=11.5pc]{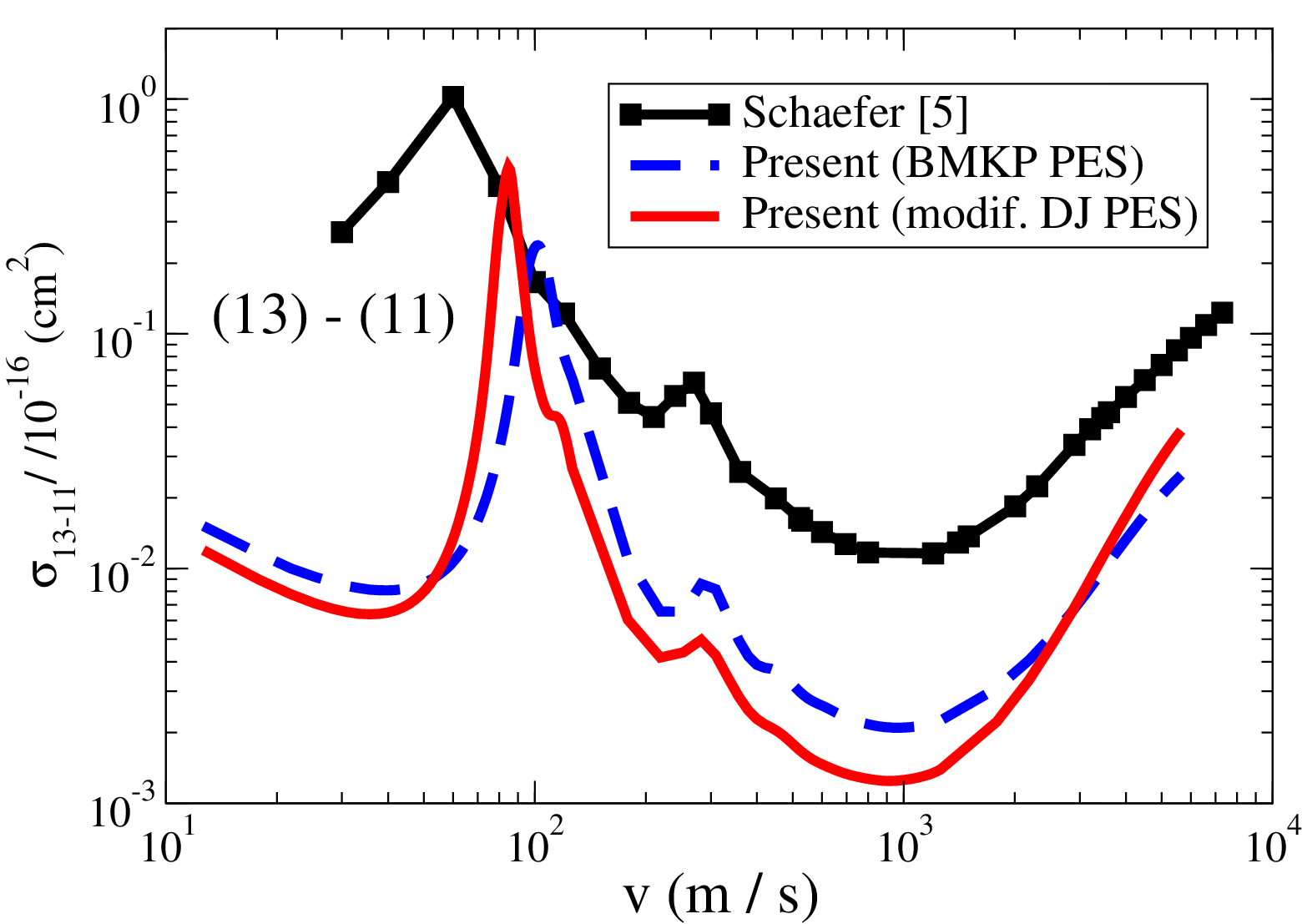}   
\caption{(Color online) Total cross sections for transition (02) $\to$ (20) 
and (13) $\to$ (11), i.e. HD(0)+H$_2$(2) $\rightarrow$ HD(2)+H$_2$(0) (upper 
panel) and 
HD(1)+H$_2$(3) $\rightarrow$ HD(1)+H$_2$(1) (lower panel) {
for different velocities $v$}. Present calculations with the BMKP \cite{schaefer} and modified 
DJ PESs are compared with those of Ref.~\onlinecite{booth}.} \label{fig4}  
\end{center} \end{figure}

\begin{figure}
\begin{center}
\includegraphics[width=.92\linewidth,clip,height=11.5pc]{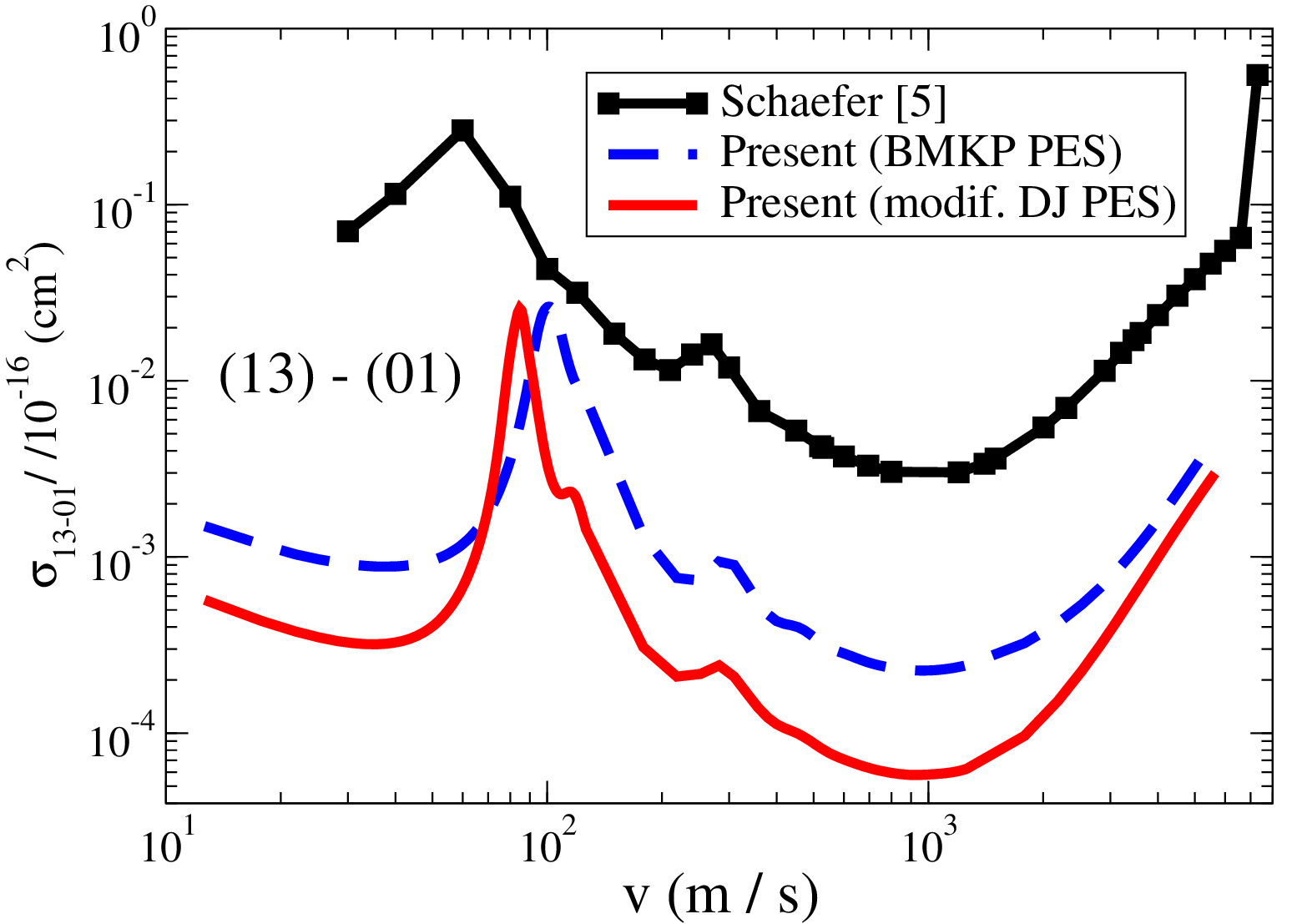}
\includegraphics[width=.92\linewidth,clip,height=11.5pc]{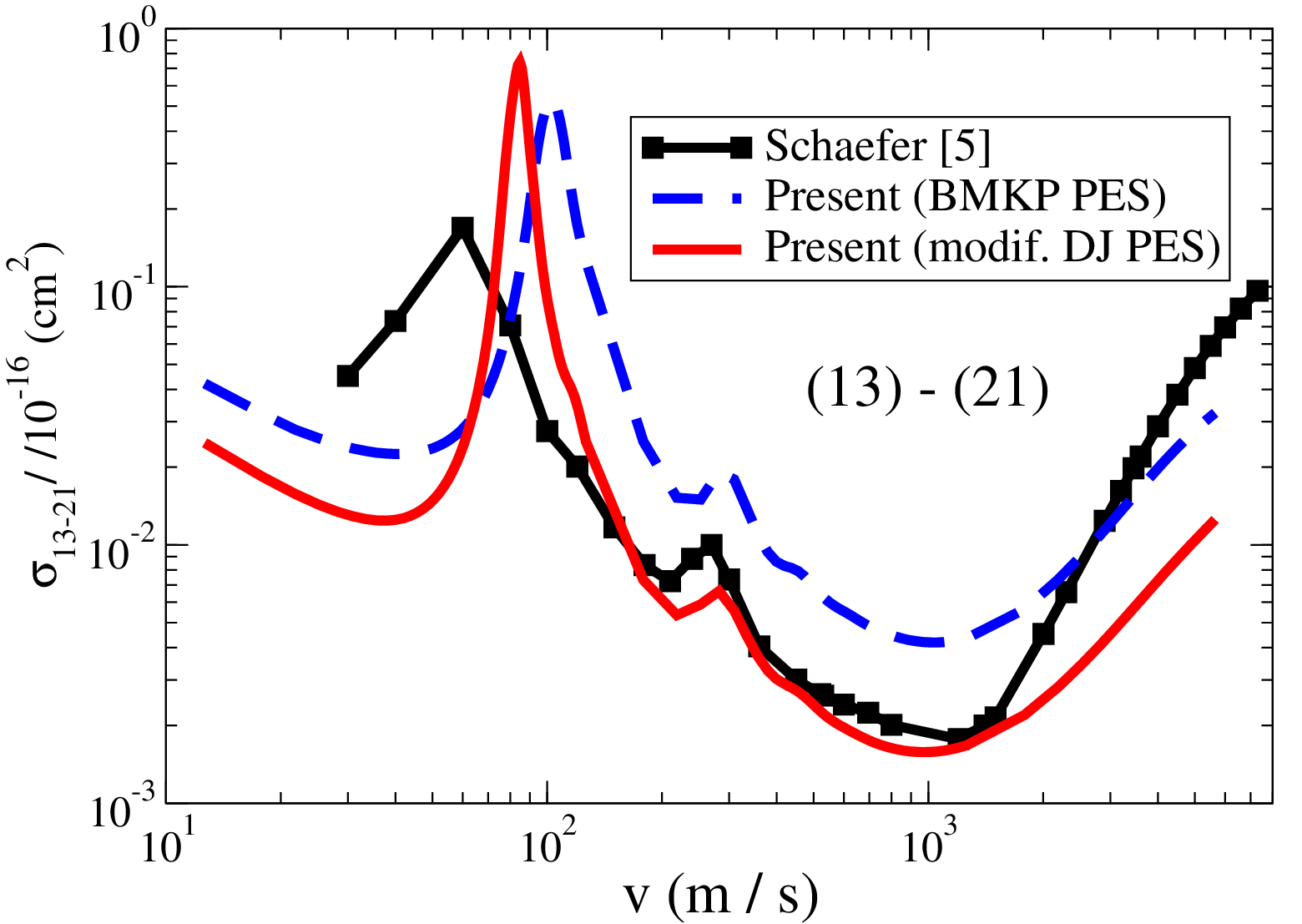}
\caption{(Color online) Same as Fig. \ref{fig4} 
for transitions (13) $\to$ (01) and (13) $\to$ (21).
}
\label{fig5}  
\end{center}
\end{figure}

\begin{figure}
\begin{center}
\includegraphics[width=.92\linewidth,clip,height=11.5pc]{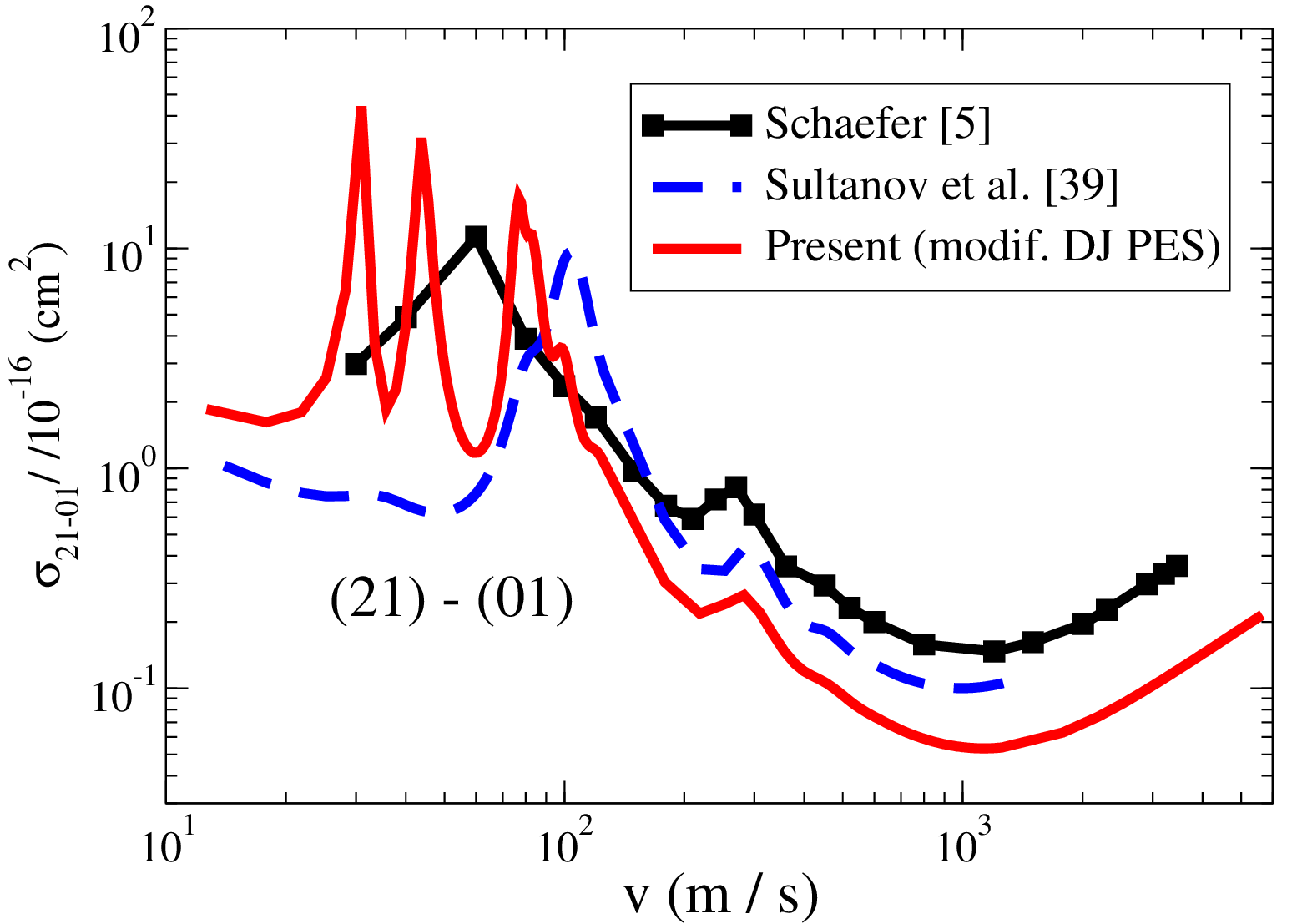}
\includegraphics[width=.92\linewidth,clip,height=11.5pc]{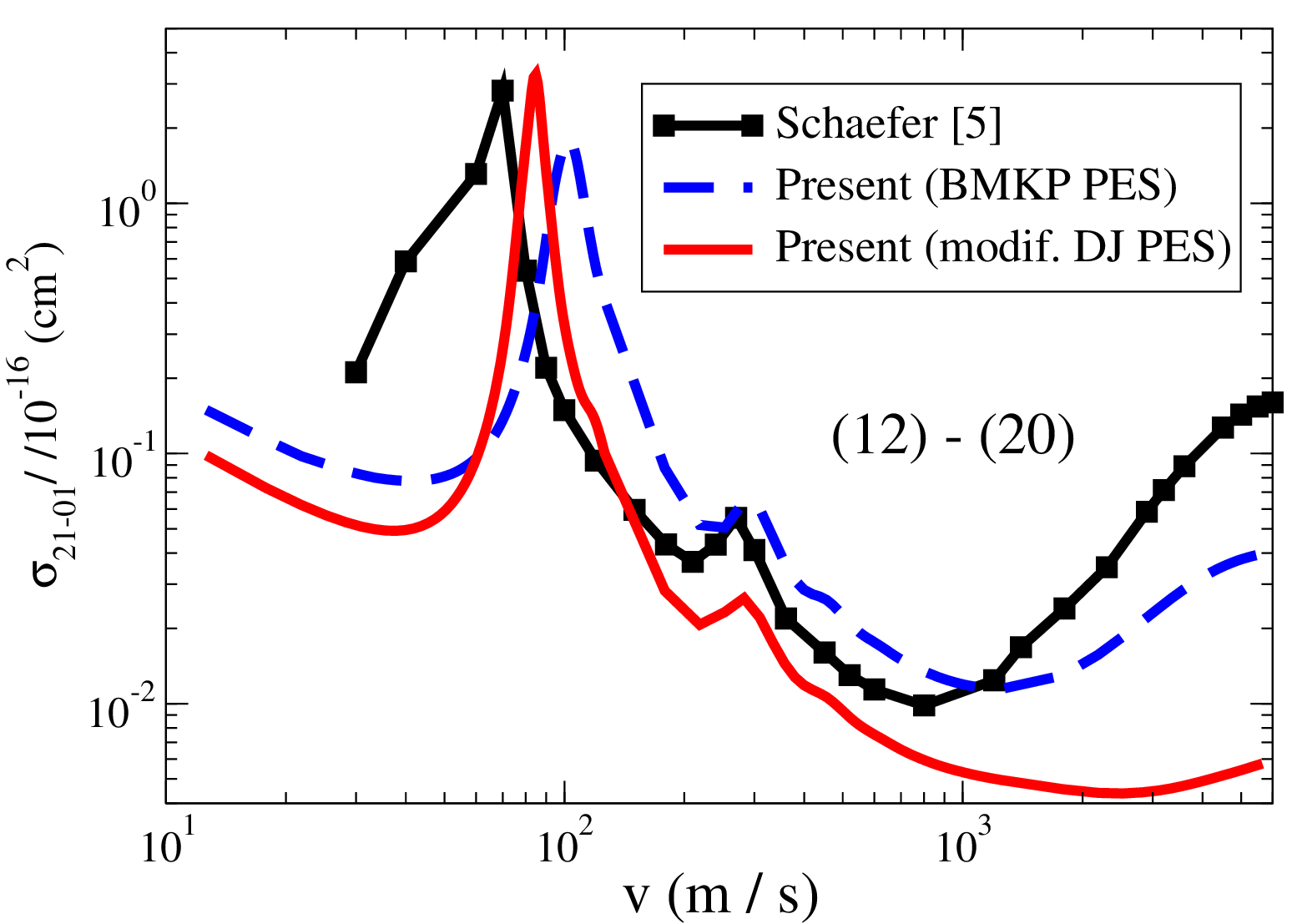}
\caption{(Color online)  Same as Fig. \ref{fig4}
for transitions (21) $\to$  (01) and (12) $\to$  (20).
}
\label{fig6}  
\end{center}
\end{figure}

\begin{figure}
\begin{center}
\includegraphics[width=.92\linewidth,clip,height=11.5pc]{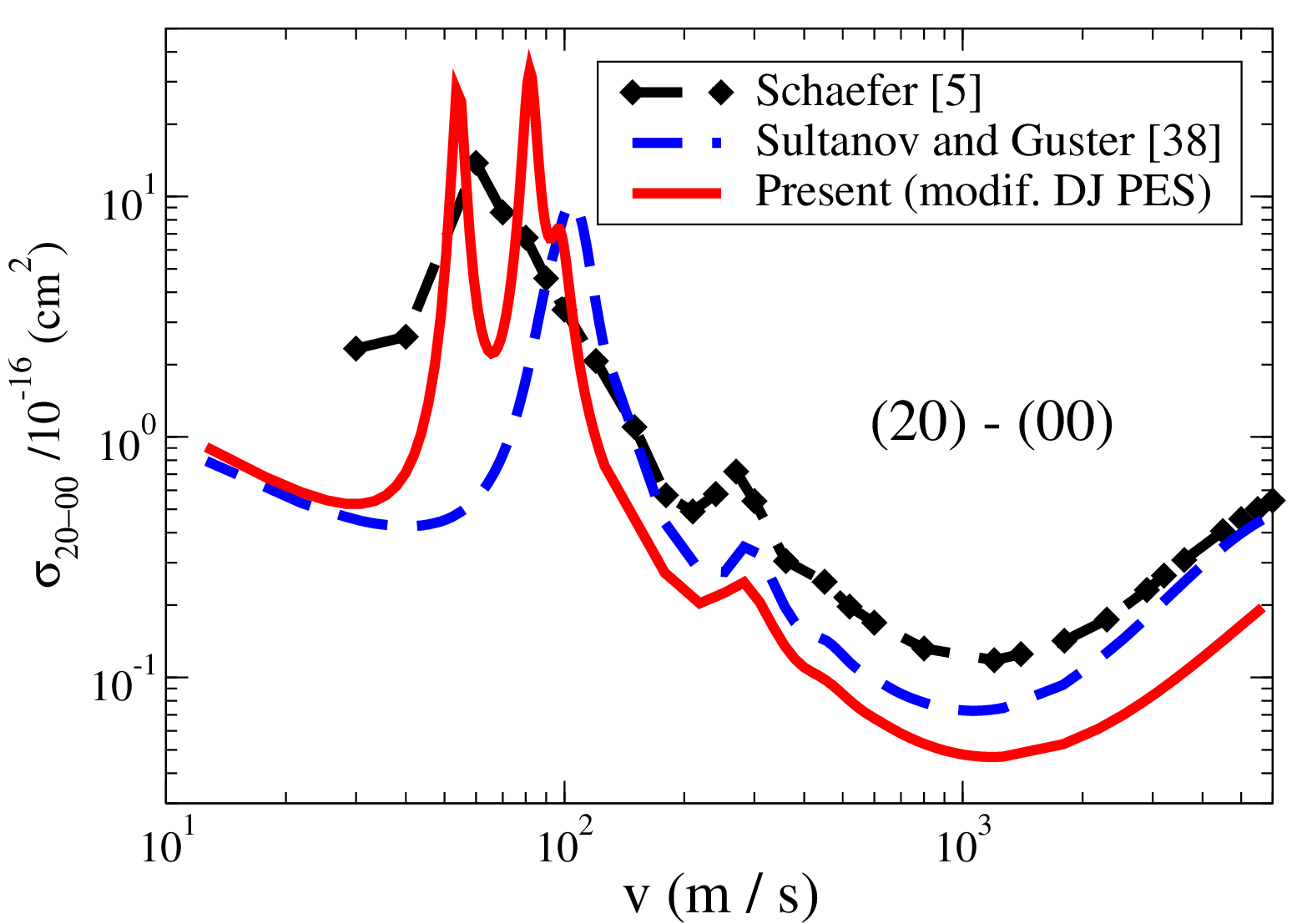}
\includegraphics[width=.92\linewidth,clip,height=11.5pc]{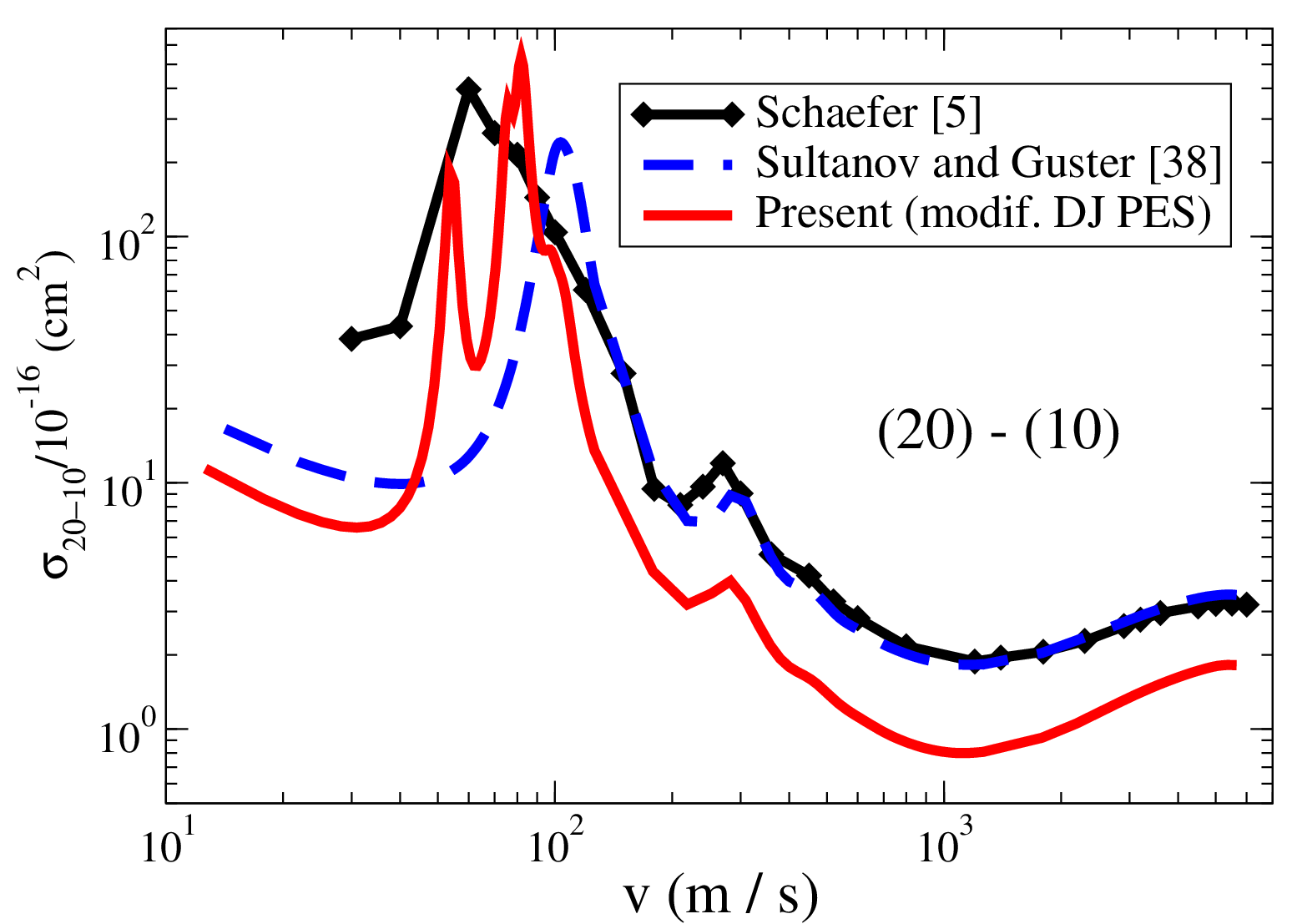}
\caption{(Color online)  Same as Fig. \ref{fig4}
for transitions (20) $\to$  (00) and (20) $\to$  (10).
}
\label{fig7} 
\end{center}
\end{figure}

\begin{figure} [!t]
\begin{center}
\includegraphics[width=.92\linewidth,clip,height=11.5pc]{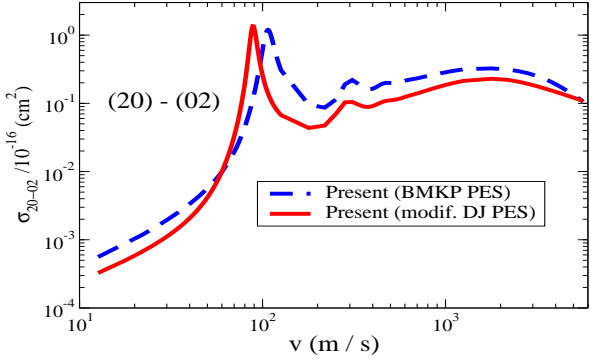}
\vspace{2mm}
\caption{(Color online)
Same as Fig. \ref{fig4} 
for transition (20) $\to$ (02).}
\label{fig8} 
\end{center}
\end{figure}

In Figs.  \ref{fig4} to \ref{fig8} we show these results for a few 
selected state-to-state HD+$o$-/$p$-H$_2$ integral cross sections. The 
results have been calculated with two different potentials, specifically, 
with the BMKP PES \cite{booth} of Sec. \ref{IIB} and the modified DJ PES
of Sec. \ref{IIC}.  When 
possible we compare our results with existing previous calculations for 
the state-selected total cross sections and thermal rate coefficients. 
It is necessary to mention that the original DJ surface 
\cite{dj} cannot be correctly used for the non-elastic or transition 
channel calculations, i.e. for the rotational state transitions in an 
HD+H$_2$ collision. If the original, unmodified DJ potential is 
applied one can get extremely low numbers $(\sim 10^{-35})$ for the 
HD+H$_2$ rotational state-to-state probabilities and cross sections. 
That is why it was necessary to undertake the geometrical modifications 
to the DJ surface described in Sec. \ref{IIC}.
We obtained a fairly good agreement between our calculations with the 
use of the two PESs. Besides some qualitative differences in the 
state-resolved total cross sections the overall behavior of 
the cross section {
$\sigma_{j_1j_2\rightarrow j_1'j_2'}(v)$ was found identical, where 
$v$ is the relative velocity of the two molecules. 
However we  
obtained substantial differences between our 
$\sigma_{j_1j_2\rightarrow j_1'j_2'}(v)$  cross sections
and corresponding data from Ref.~\onlinecite{schaefer}.}

In  Fig. \ref{fig4} we plot the total cross sections for the rotational
transitions (02) - (20) and (13) - (11). 
The notation of the rotational quantum numbers can be
understood by comparison to the following equations:
\begin{eqnarray}
\mbox{HD}(0)+\mbox{H}_2(2)&\rightarrow&  \mbox{HD}(2)+\mbox{H}_2(0), \\ 
\mbox{HD}(1)+\mbox{H}_2(3)&\rightarrow& \mbox{HD}(1)+\mbox{H}_2(1),
\end{eqnarray}
{where the numbers in parenthesis denote rotational quantum 
numbers $j_1$, $j_2$ etc.}.
On the upper plot it is seen that while the two results of this 
study are in fairly good agreement 
between each other we obtain substantial
disagreements with the result of Ref.~\onlinecite{schaefer}.
The same is true in the lower plot, although
the behavior of these cross sections has a common character.

In Fig. \ref{fig5} we show the cross sections for rotational transitions 
 in HD+H$_2$ for the BMKP and the modified DJ PESs
for the processes 
$\mbox{HD}(1)+\mbox{H}_2(3)\rightarrow\mbox{HD}(0)+\mbox{H}_2(1),$
and 
$\mbox{HD}(1)+\mbox{H}_2(3)\rightarrow \mbox{HD}(2)+\mbox{H}_2(1).$
The corresponding results from Ref.~\onlinecite{schaefer} are also shown.
We again obtain a fairly good agreement between our results computed 
with the BMKP and the modified DJ PESs,
however, differing significantly from the corresponding 
Schaefer result \cite{schaefer}.
Unfortunately we cannot compare these cross sections with the results 
of the calculation by Flower \cite{flower99}, in which a
different H$_2$-H$_2$ potential from \cite{schwenke}  was used. This is because
Flower's data includes results mostly for the thermal rate coefficients. 
However, within the next subsection
in Table II we compare our few
selected rotational state-resolved thermal 
rate coefficients with the corresponding results from 
Refs.~\onlinecite{schaefer,flower99}. 

From the astrophysical point of view, perhaps, one needs only precise 
rotational and in some rare cases vibrational state-to-state thermal 
rate coefficients {
$k_{j_1j_2\rightarrow j_1'j_2'}(T)$ in H$_2$+H$_2$, HD+H$_2$ 
etc.} These quantities are less 
sensitive to interaction potentials. However, the overall behavior of 
all possible {state-selected} cross sections should be very 
important for calculation of the thermal rates as seen in Fig. 
\ref{fig4}. It appears that the cross sections are much more sensitive 
to the PESs used in the calculations. Hence it is useful and even 
probably important in some specific cases to compare the cross sections 
from various calculations where different PESs have been used.

Further, Figs. \ref{fig6} and \ref{fig7} exhibit our results for the 
state-to-state rotational cross sections in transitions 
$\mbox{HD}(2)+\mbox{H}_2(1)\rightarrow \mbox{HD}(0)+\mbox{H}_2(1)$
and $\mbox{HD}(1)+\mbox{H}_2(2)\rightarrow \mbox{HD}(2)+\mbox{H}_2(0)$
 and  in transitions $\mbox{HD}(2)+\mbox{H}_2(0) 
\rightarrow \mbox{HD}(0)+\mbox{H}_2(0)$
and $\mbox{HD}(2)+\mbox{H}_2(0)\rightarrow \mbox{HD}(1)+\mbox{H}_2(0)$,
 respectively. 
We obtained fairly good agreement between our own results.  
Additionally, in these rotational transitions fairly good agreement 
with the corresponding cross sections from Ref.~\onlinecite{schaefer} has also been obtained.
Finally, Fig. \ref{fig8} shows our integral cross section for the transition
$\mbox{HD}(0)+\mbox{H}_2(2)\rightarrow \mbox{HD}(2)+\mbox{H}_2(0)$.
This process is also interesting, because the transition occurs in the two molecules 
simultaneously. We would like to name such processes as 
double-transition processes.

\begin{table*}
\label{TII}

\caption{Results for three selected state-to-state rotational thermal 
rate coefficients 
$k_{j_1j_2\rightarrow j_1'j_2'}(T)$ 
cm$^3$ s$^{-1}$ at various 
low temperatures $T$ (K) in the $p$-H$_2(\alpha)$ + HD$(\beta)$ $\rightarrow$ 
$p$-H$_2(\alpha')$ + HD$(\beta')$ collisions. Calculations with different PESs: 
the original BMKP PES \cite{booth} and the new modified DJ potential 
from this work. The corresponding older data from other authors 
\cite{schaefer,flower99} are also included in this table. Numbers in 
parentheses are powers of 10.}
\vspace{3mm}
\centering
\label{table:2}
\begin{tabular}{lcccccccccccccccccccccccc}
\hline
\hline
$T$ (K) & & \multicolumn{12}{c}{Rotational Thermal Rate 
Coefficients: $k_{j_1j_2\rightarrow j'_1j'_2}(T)$ cm$^3$ s$^{-1}$}\\
\hline
&\multicolumn{4}{c}{(02)$\rightarrow$(20)}& &\multicolumn{4}{c}{(12)$\rightarrow$(20)}&
&\multicolumn{4}{c}{(20)$\rightarrow$(02)}&\\ 
\hline
& BMKP  & Mod. DJ   &Ref.~\onlinecite{flower99}&Ref.~\onlinecite{schaefer}&   & BMKP  & Mod. DJ &Ref.~\onlinecite{flower99}&Ref.~\onlinecite{schaefer} &   &
  BMKP  & Mod. DJ   &Ref.~\onlinecite{flower99}&Ref.~\onlinecite{schaefer}\\ 
\hline
%
%
10 & 1.15(-11) &  6.38(-12)  &          & 7.21(-15) &  & 1.50(-13) & 6.35(-14) &           & 9.32(-14) &
   & 1.06(-17) &  3.43(-18)  &          & 2.53(-20) &\\
20 & 9.57(-12) &  5.51(-12)  &          & 6.72(-15) &  & 1.27(-13) & 5.43(-14) &           & 8.62(-14) &
   & 9.18(-14) &  4.11(-15)  &          & 1.26(-17) &\\
30 & 9.05(-12) &  5.40(-12)  &          & 7.40(-15) &  & 1.25(-13) & 5.34(-14) &           & 9.53(-14) &
   & 8.80(-14) &  4.48(-14)  &          & 1.13(-16) &\\
50 & 8.81(-12) &  5.51(-12)  & 9.3(-13) & 1.09(-14) &  & 1.34(-13) & 5.64(-14) & 7.9(-14)  & 1.35(-13) &
   & 5.47(-13) &  3.14(-13)  & 7.5(-14) & 8.86(-16) &\\
70 & 8.83(-12) &  5.68(-12)  &          & 1.70(-14) &  & 1.52(-13) & 6.06(-14) &           & 1.89(-13) &
   & 1.21(-12) &  7.38(-13)  &          & 2.82(-15) &\\
100& 8.93(-12) &  5.90(-12)  & 9.9(-13) & 3.21(-14) &  & 1.85(-13) & 6.72(-14) & 1.6(-13)  & 2.86(-13) &
   & 2.23(-12) &  1.42(-12)  & 2.8(-13) & 9.15(-15) &\\
200& 9.17(-12) &  6.29(-12)  & 1.1(-12) & 1.44(-13) &  & 3.25(-13) & 8.78(-14) & 4.3(-13)  & 7.31(-13) &
   & 4.58(-12) &  3.10(-12)  & 6.0(-13) & 7.69(-14) &\\
300& 9.19(-12) &  6.43(-12)  &          & 3.46(-13) &  & 4.77(-13) & 1.06(-13) &           & 1.28(-12) &
   & 5.78(-12) &  4.02(-12)  &          & 2.28(-13) &\\
\hline
\hline
\end{tabular}
\end{table*}

\subsection{Thermal rate coefficients}

\label{IIIC}

In Table II our rotational thermal rate coefficients 
$k_{j_1j_2\rightarrow j_1'j_2}(T)$
are presented. In this paper we choose only three rotational 
double-transitions in the 
collision: (02)-(20), (12)-(20), and (20)-(02). 
These results were obtained from corresponding
state-resolved integral cross sections 
$\sigma_{j_1j_2\rightarrow j_1'j_2}(\epsilon)$ with the use of the expression (\ref{eq:kT}).
In a previous  
study \cite{flower99},
Flower compared his rotational transition thermal rate coefficients 
with the corresponding Schaefer data \cite{schaefer} and 
found substantial differences between his results and results from 
Ref.~\onlinecite{schaefer} even at
high temperatures.
The point is that in these processes the transition occurs in the two 
molecules simultaneously. In such 
rotational transitions the probabilities and cross sections should be 
very sensitive to the interaction potential. Perhaps, because of this 
reason the results of Refs.~\onlinecite{schaefer,flower99} differ 
so dramatically for transitions like (02) - (20). In fact, we also 
obtained substantial differences between our thermal rates and with 
corresponding results from Refs.~\onlinecite{schaefer,flower99}. Particularly 
the large differences were detected at the lower temperature regime.

In Table II we show our thermal rate coefficients 
{ $k_{j_1j_2\rightarrow j_1'j_2'}(T)$} for the processes
$\mbox{HD}(0)+\mbox{H}_2(2)\rightarrow \mbox{HD}(2)+\mbox{H}_2(0),$
and 
$\mbox{HD}(1)+\mbox{H}_2(2)\rightarrow \mbox{HD}(2)+\mbox{H}_2(0),$
and those for 
$\mbox{HD}(2)+\mbox{H}_2(0)\rightarrow \mbox{HD}(0)+\mbox{H}_2(2).$
We present our results calculated with the BMKP \cite{booth} and with 
the modified DJ \cite{dj} PESs. As before, our results computed 
with these two potentials are close to each other. However, one can see 
that our results and results from Refs.~\onlinecite{schaefer,flower99} differ 
significantly. This happens particularly at low temperatures, for 
instance from 10 K to 50 K. Finally in this section, for the process 
$\mbox{HD}(2)+\mbox{H}_2(0)\rightarrow \mbox{HD}(0)+\mbox{H}_2(2)$
the difference between our $k_{j_1j_2 \rightarrow j_1'j_2'}(T)$ and 
the results from Ref.~\onlinecite{schaefer} is about 3 orders of 
magnitude at T=10 K. The reason of this substantial deviation is not clear,
although it might be a result of using different HD+H$_2$ potentials
in the current computation and in Refs.~\onlinecite{schaefer,flower99}, where
the authors also used two different PESs and different quantum-mechanical
methods. However, we would like to point out here that our results computed
with two newer PESs \cite{dj,booth} are relatively close to each other.

\subsection{Application of the detailed balance principle}

\label{IIID}

By using the time reversibility (reciprocity) principle one
can obtain the detailed balance equation for the direct and reverse
energy transfer processes or reactions\cite{landau}. 
In the case of the
inelastic scattering $a + b \rightleftharpoons a^{\prime} + b^{\prime}$ 
the detailed balance principle\cite{landau}
relates the direct $a+b$ and the reverse $a'+b'$ processes and their cross sections
$\sigma^{ab}_{j_aj_b\rightarrow j'_aj'_b}$ and $\sigma^{a'b'}_{j'_aj'_b\rightarrow j_aj_b}$:
\begin{eqnarray}
(2j_a+1)(2j_b+1)p^2_{a\rightarrow b}\sigma^{ab}_{j_aj_b\rightarrow j'_aj'_b}(E) & = &
(2j'_a+1)\nonumber \\
\times (2j'_b+1) p^2_{b\rightarrow a}\sigma^{a'b'}_{j'_aj'_b\rightarrow j_aj_b}(E).
\label{eq:db1}
\end{eqnarray}
Here, $E$ is the total energy,
$j_{a(b)}$ and $j'_{a(b)}$ are the initial and final rotational quantum numbers,
$p^2_{a(a')\rightarrow b(b')}$ are the initial and final relative momenta between $a$ and
$b$ species, $\sigma^{ab}_{j_aj_b\rightarrow j'_aj'_b}(E)$ and 
$\sigma^{a'b'}_{j'_aj'_b\rightarrow j_aj_b}(E)$ are direct and reverse integral cross sections
respectively. The same type of the relationship
can be obtained for the thermal rate coefficients $k_{j_1j_2\rightarrow j'_1j'_2}(T)$. 
Let us rewrite Eq. (\ref{eq:kT}) in the terms of the total energy $E$. Considering
that the relative kinetic energy $\epsilon$ between $a$ and $b$ is Eq. (\ref{eq:ekin}),
we compute the total energy from the lowest possible level between two channels.
If it is associated with the second channel $a'+b'$, i.e. $j'_1j'_2$ pair,
the formula (\ref{eq:kT}) becomes:
\begin{eqnarray}
k^{ab}_{j_1j_2\rightarrow j'_1j'_2}(T) &=& \frac{1}{(k_BT)^2} \sqrt{ \frac{8k_B T}{\pi M_{12}}}
\int_{0}^{\infty}
\sigma_{j_1j_2\rightarrow j'_1j'_2}(E)\nonumber \\
&\times &
\frac{p^2_{ab}}       
{M_{12}}
e^{-(E-\Delta E) / k_BT}dE.
\label{eq:kT2}
\end{eqnarray}
Here, $\Delta E = [B_1j_1(j_1+1) + B_2j_2(j_2+1)] - [B_1j'_1(j'_1+1) + B_2j'_2(j'_2+1)]$
is the energy gap between the direct and reverse channels, and
$\epsilon = p^2_{ab}/M_{12}$ is the kinetic energy. Now,
for the reverse channel the thermal rate coefficient is:
\begin{eqnarray}
k^{a'b'}_{j'_1j'_2\rightarrow j_1j_2}(T) &=& \frac{1}{(k_BT)^2} \sqrt{ \frac{8k_B T}{\pi M_{12}}}
\int_{0}^{\infty}
\sigma_{j'_1j'_2\rightarrow j_1j_2}(E)\nonumber \\
&\times & 
\frac{p^2_{a'b'}}      
{M_{12}}
e^{-E/k_BT}dE,
\label{eq:kT3}
\end{eqnarray}
Comparing Eqs. (\ref{eq:kT2}) and  (\ref{eq:kT3}) and taking into account Eq.
(\ref{eq:db1}) we obtain the detailed balance formula for the thermal rate coefficients:
\begin{eqnarray}
(2j_1+1)(2j_2+1) k^{ab}_{j_1j_2\rightarrow j'_1j'_2}(T)
 &=&
(2j'_1+1)(2j'_2+1)\nonumber \\ 
\times k^{a'b'}_{j'_1j'_2\rightarrow j_1j_2}(T) e^{\Delta E/k_BT}.
\label{eq:db2}
\end{eqnarray}
The ratio 
$R_{j_1j_2\rightleftharpoons j'_1j'_2}(T) = k^{ab}_{j_1j_2\rightarrow j'_1j'_2}(T) / k^{a'b'}_{j'_1j'_2\rightarrow j_1j_2}(T)$
is proportional to an exponent with the argument depending on $\Delta E$.
In this work we computed two direct-reverse processes in the HD + $p$-H$_2$
collisions, specifically:
\begin{equation}
\mbox{HD}(0)+\mbox{H}_2(2)\rightleftharpoons \mbox{HD}(2)+\mbox{H}_2(0),
\label{eq:0220}
\end{equation}
with $\Delta E=96.6$ cm$^{-1}$.
It would be useful to check how well the computed
thermal rates (\ref{eq:kT}) in this work 
satisfy the detailed balance equation (\ref{eq:db2}).
\begin{figure} [!t]
\begin{center}
\includegraphics[width=.92\linewidth,clip,height=11.5pc]{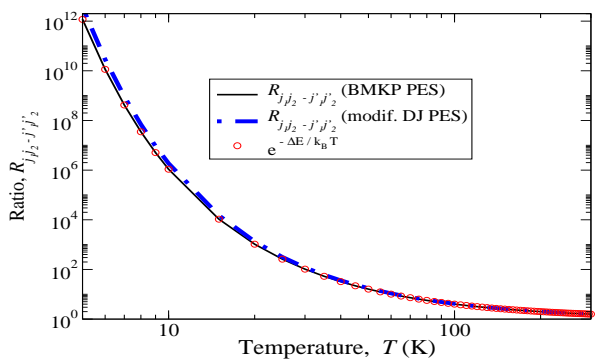}
\vspace{2mm}
\caption{(Color online) The ratio $R_{j_1j_2\rightleftharpoons j'_1j'_2}(T)$
for the processes (\ref{eq:0220}). Computation with the use of both potentials:
the BMKP and the modified DJ PESs. The open circles are the values of the exponential function from
the right side of Eq. ({\ref{eq:db2}}).}
\label{fig9}
\end{center}
\end{figure}
Fig. 9 represents these results. It is shown
that all results are in a satisfactory agreement with each other.

\section{Conclusion}

\label{IV}

In this work
a rotational method has been applied  for the transformation of the 4-dimensional
rigid monomer model H$_2$-H$_2$ PES 
of Ref.~\onlinecite{dj} to the non-symmetrical potential appropriate for calculations of the HD+H$_2$ collisions. 
Different low energy
elastic and state-selected inelastic cross sections as well as the 
thermal rate coefficients for HD+H$_2$ have been computed
and compared with previous calculations, where available.  The rotational 
energy transfer in HD+H$_2$ is of importance for the 
thermodynamics of the ISM\cite{dalg}.
By now few and rather conflicting results are available 
for the low energy HD+H$_2$ rotational energy transfer, see for example
\cite{my22} and references therein. The BMKP PES \cite{booth} has been 
already applied to HD+H$_2$ \cite{my2,my22}. However, this PES may 
have failures. The fact was mentioned in Refs.~\onlinecite{my1,lee} and in \cite{my11}.
Therefore in this paper a new attempt has been undertaken to carry out
alternative computational methods for HD+H$_2$ collision.
%
%
In case of the BMKP and DJ PESs 
the necessary steps       for each potential have been described in 
Secs. \ref{IIB} and  \ref{IIC} and also in Ref.~\onlinecite{my2}. In the case of the BMKP potential, which is a full 
six-dimensional surface \cite{booth}, the transformation from 
H$_2$-H$_2$ to the H$_2$-HD system was done by shifting the center of 
mass in one H$_2$ molecule to the center of mass of the HD molecule.  
Because the DJ PES has been formulated for the rigid monomer rotor model, 
the transformation methodology was more complicated. Simply, the $\vec R_1$  and $\vec R_2$
coordinates are not available in this case, they have fixed lengths.
In this paper the transformation has been accomplished 
by rotation of the space fixed $OXYZ$ coordinate system, i.e. by the 
redirection of the $\vec R_3$ vector. The new vector $\vec R'_3$ 
connects the center of masses of the H$_2$ and the HD molecules as shown 
in the Fig. \ref{fig2}. This procedure obtains new coordinate angles for 
the Jacobi vectors $\vec R_1$, $\vec R_2$, and $\vec R_3$, and a new PES as in Eq. (\ref{eq:transfor}).
New experiments that measure the state-to-state rotational cross sections in the HD+$o$-/$p$-H$_2$ 
collisions at low temperatures are needed. Thereafter 
theoreticians and astrophysicists would be able to compare
computational results with available experimental data. This type of 
work was recently accomplished, for instance, for the $para$-H$_2$+H$_2$ 
collision \cite{montero05}. Here it would be useful to mention other
contributions on hydrogen-hydrogen collisions\cite{kusukabe,meyer,reddy}. 
Additionally,
with the use of new HD+H$_2$ results for the thermal rate coefficients
one could carry out new computation of the HD-cooling function mentioned in the introduction\cite{dalg}.

In conclusion, another interesting system worth mentioning is HD+HD. For 
this collision there are relatively old experimental state-to-state 
rotational probabilities for a few selected states \cite{hdhd}. These 
old data can be useful in comparisons with the computational results 
obtained with different H$_4$ PESs: such as the available DJ and the BMKP PESs or
some relatively new potentials, for example from works\cite{belof,hinde08}.
In  the case of the DJ surface it would be possible to again apply the 
rotation procedure of the $OXYZ$ coordinate 
system as performed in this paper.

\acknowledgments
This work was supported by Office of Sponsored Programs (OSP) and by Internal Grant Program of
St. Cloud State University, and CNPq and FAPESP of Brazil.

\end{document}